# Self-calibrating Intelligent OCT-SLO System

Mayank Goswami

*Divyadrishti Imaging Lab, Department of Physics, IIT Roorkee, Roorkee- 247667, India*

**mayank.goswami@iitr.ac.in**

## Abstract:

A unique sample-independent 3D self-calibration methodology is tested on a unique optical coherence tomography and multi-spectral scanning laser ophthalmoscope (OCT-SLO) hybrid system. Operators' visual cognition is replaced by computer vision using the proposed novel fully automatic AI-driven system design. Sample-specific automatic contrast adjustment/focusing of the beam is achived on the pre-instructed region of interest. The AI model deducts infrared, fluorescence, and visual spectrum optical alignment by estimating pre-instructed features quantitatively. The tested approach, however, is flexible enough to utilize any apt AI model.

Relative comparison with classical signal-to-noise-driven automation is shown to be ~200% inferior and ~130% slower than the AI-driven approach. The best spatial resolution of the system is found to be: (a) 2.41 microns in glass bead eye phantom, $0.76 \pm 0.46$ microns in the mouse retina in the axial direction, and (b) better than 228-line pair per millimeter (lp/mm) or 2 microns for all three spectrums, i.e., 488 nm, 840 nm, and 520-550 nm emission in coronal/frontal/x-y plane.

Intelligent automation reduces the possibility of developing cold cataracts (especially in mouse imaging) and patient-associated discomfort due to delay during manual alignment by facilitating easy handling for swift ocular imaging and better accuracy. The automatic novel tabletop compact system provides *true* functional 3D images in three different spectrums for dynamic sample profiles. This is especially useful for photodynamic imaging treatment.

## 1. Introduction

Optical Coherence Tomography (OCT) is an advanced functional imaging technique used for time-dependent 3D topographical measurement and visualization. The only requirement is the availability of a transparent (to a certain wavelength) path till the intended Region of Interest (ROI). This requirement defines its possible utility [1]. Despite the limitation, this technique has found its use in clinical diagnostics, treatment interventions in the fields of optometry & ophthalmology[2], dermatology [3], subsurface imaging of roots of skin lesions, superficial coronary artery plaques diagnostics [4], and several other non-clinical applications[5].

The OCT is hybridized with X-ray Computerized Tomography (CT) or with Adaptive Optics Scanning Laser Ophthalmoscope (AO-SLO) to integrate the advantages of the respective techniques[6], [7]. For example, OCT integration with SLO improves the ease of handling the OCT system. The clinical interpretation of anatomical structure improves by aligning/colocalizing the SLO enface images with averaged 3D OCT images in an axial direction [8], [9]. These multimodal/multi-spectral non-invasive imaging tools are used to study (a) novel diagnostics mechanisms, (b) the progression of disease models, (b) drug delivery mechanisms, and (c) imaging-assisted treatment protocols in-vivo.

Optics-based functional imaging systems may require hardware alterations or optical realignment to cater to the need on a case-to-case basis. Inventions such as handheld OCT for pediatricians, OCT for Owl eye imaging, chip-based OCT to integrate with endoscopy, etc., are a few of the examples. In summary, the main goal remains to miniaturize the overall system size, increase the ease of handling, and/or improve the overall technical specification, etc.[10], [11], [12]. High spatial and/or contrast resolution, broad field of view (FOV), accurate calibration, and high scanning speed are a few of the specifications that commercial suppliers might improve upon to claim the title of provider of state-of-the-art systems. The following subsections briefly discuss factors affecting spectrum-specific resolution and calibration accuracy.

## 1.1. Aberration and resolution

Several techniques are proposed to enhance an optical imaging system's possible resolution (under the diffraction limit). The resolution can be increased by using better detectors/cameras, optics, and sample-specific apt alignment, providing minimum optical aberrations and dispersion.

Several hardware designs and analytical methods are reported to achieve these goals. Expensive wavefront sensors and deformable mirrors are used to minimize the aberrations by physically sensing and accordingly adjusting the wavefront [13], [14], [15]. The OCT and sensorless AO systems can also manage the same by applying soft numerical aberration correction techniques during post-processing. Numerical estimation and compensation of (a) an aberrated wavefront or (b) an optimal complex pupil phase profile are two main computational correction techniques. Coefficients of the Zernike polynomials are optimized using classical or backpropagation algorithms to maximize image sharpness metrics[16].

## 1.2. Focusing, Dispersion, contrast, and enface resolution

Loss of axial spatial resolution happens due to dispersion mismatch and sub-optimal focusing. Manually replacing and testing the dioptre of different lenses, if the ROI profile changes, is slow and impractical. Beam steering and the use of a variable focus lens (VFL) are a few possibilities when focus in the same plane needs to be changed. Achromatic VFL is needed if multi-spectral imaging is performed. An innovative solution is to use a variable voltage liquid lens[17], [18]. Varying the lens voltage, however, may affect the axial focus of light. Focusing on the enface plane may strain the operator's ability to visualize the focus in the axial direction in 2D images at the display. It strains the processing hardware because RAW data to 3D data averaging is required. It might delay the real-time display, let alone the issues related to observing any visual difference while varying the associated parameters.

Dispersion mismatch can be reduced by digitally suppressing the frequency-dependent nonlinear phase, mainly second-order group velocity dispersion and third-order asymmetric distortion components during the post-processing stage or convoluting the depth scan signal with a depth variant kernel, a posteriori [19], [20]. It can also be balanced by introducing apt transparent films[21].

The OCT system comprises the Mach-Zehnder interferometer[22], [23]. In the case of biomedical imaging, sample sizes may vary. The operator may be expected to adjust one of the arms (preferably the reference arm mounted with a mirror (RM)) of the Mach-Zehnder interferometer to minimize the imbalance of dispersion and thus resolve a certain depth[16], [24]. The optimal setting may depend on the globe size of the animal eye (if the axial scanning length is greater than 2mm)/type of sample, mechanical vibrations, ambient air, and temperature changes [19], [25]. Commercial systems (a packed assembly), however, may facilitate the same option by integrating this arm with a stepper motor that is semi-automated with a graphical user interface (GUI)[26]. The stepper motor attached to one of the arms allows the operator to optimize the arm length without directly touching the optics[27], [28].

Contrast resolution can be improved by adjusting the optical alignment by focusing the light under the ROI. The sample's distance from the scanning lens can be adjusted for the same purpose. The reference arm position and/or axial distance of the sample to the scan lens can be adjusted simultaneously to match the optical path from reflections within the ROI (for example, retina). It also allows ROI to be brought to the display images' center with appropriate brightness/power deposition. Variable curvature liquid lenses are used for contrast/aberrations adjustment to reduce the speckle noise[23], [29].

## 1.3. Calibration and Spatial resolution for functional imaging

Standard calibration phantoms such as USAF typically have prefabricated structures with several known dimensions. The average calibrating factor is the mean ratio of the number of pixels covering several structures (with different dimensions) by their actual size (physical true width and length provided by the manufacturer). This calibration factor can be used to estimate the true size of structures in other samples when the same imaging system is used.

Spatial resolution is estimated by imaging the finest structure. The two finest structures available near to each other, if they are smaller than the system can image, will look like they merged into each other and are relatively bigger than their actual physical dimensions. Using these structures (due to some manual or automatic matching process) will result in an erroneous calibration. Thus, accurately knowing the resolution of the system before calibration is essential for one-to-one correlation. Typically, a system with finer spatial resolution (than the dimension of the structures in the sample) is used for true functional imaging.

It is assumed that the path's refractive index (RI) and the ROI are the same for the standard Phantom and the sample (to be imaged) with unknown dimensions. This assumption may not be correct because the path and ROI index may not be the same for the Phantom and sample. Counting the pixel under ROI is an image processing step, and if done manually, it may further impart error. A systems resolution may, thus, depend on the RI of the path and ROI itself.

For biological samples, anatomical ROI/structural size (for example, globe size) may not be the same for all [30], [31], [32]. Biological samples such as the eye may require sacrificing the animal, and anatomical structure may collapse in the absence of natural interstitial (osmotic, hydrostatic, and/or lymphatic) pressure, thus not maintaining the true dimensions during histopathological analysis [33], [34]. Knowing their true physical dimension may not be thus possible. For human ocular imaging, this data may not be available.

Thus, a calibration phantom with RI similar to the sample (to be imaged) may be needed or measurement of true physical dimensions must be done for a tagged structure in the same sample. A sample (to be imaged) itself may be used for calibration, curbing the need for relative comparison with standard Phantom.

Typical Focus-tracking techniques implemented for optical coherence microscopy (OCM) are used to improve the axial resolution; however, they offer limited magnification. Autofocusing techniques either use ultrasound or infrared to estimate the distance of ROI to translate the lens at an apt position or iteratively vary the lens position for a sharp image. The variation of lens position either obtained manually or using a linear scanning galvanometer, is a slow and less accurate method [35].

*The discussion above highlights the need for a sample-dependent calibration method, which can be applied to any sample to estimate the sample-specific calibration factor without needing manual intervention/automatically.* There, however, are no phantoms to calibrate along the axial directions.

## 1.4. Advantage of automation

A multi-spectral functional imaging system has several optomechanical components running together. The operator may have to find optimal settings simultaneously. Common optimal operating parameters and alignment settings for sample-dependent self-calibration cannot be achieved in a single/direct step. These mutual adjustments alter physical and electronic parameters and, thus, require a certain amount of time. Instead, iterative adjustments while analyzing the desired features from images is a usual approach that is carried out by the operator. The exercise requires practice as the depth of focus in ROI and/or morphology may vary from animal to animal, person to person, and sample to sample. A situation of doubt about the quality may arise if only a few discrete combinations are tried, even if the operator has expert experience. Furthermore, it is challenging in real time to accurately compare and decide if the current 3D volume of the Region of Interest (ROI) is superior to what was visualized using previous settings.

The parameter variation needs to be continuous. Ideally, it should be non-manual to encompass all potential combinations of parameters, thereby eliminating the need to manually set only a few specific values within a limited timeframe. This approach aims to achieve high resolution and accuracy from the imaging system. By ensuring continuous variation and minimizing manual intervention, the automatic approach seeks to optimize the system's performance by covering a wide range of parameter settings swiftly and accurately. Finding the optimal setting and finding it fast is crucial. Every minute counts, especially in the case of animal ocular imaging. A mouse, for example, on stage under anesthesia, may develop a temporary cold cataract into vitreous fluid, disturbing the necessary transparency to the retina blocking the light [36]. Any delay while the anesthetized animal is on the stage may render its eye useless for a few hours. A trade-off between imaging time and accuracy to align must be achieved.

Visual comparison (required in semi-automatic system design) of two A/B Scans on two different settings is a qualitative approach. The operator handling more than one button (hardware or via GUI) and the anesthesia system when the mouse is on stage may experience stress and finalize sub-optimal settings. It is possible that better quality of A/B Scan may be acquired if the quantitative analysis is used in real-time. Multiple operators may add up to the cost of operation and require coordination.

Classical approach-based automation (using an analytical expression curve fitted from previously obtained data) may fall into local extremum solutions and thus require a choice of robust optimization/mathematical approaches with respect to data. It may only work for samples for which the curve has been fit. The analytical expression consistently requires calibration updates as the operational lifetime passes and may vary from system to system. Automatic tuning of only one of the factors at a time for a single spectral imaging system is provided in the literature. Signal to noise ratio maximization-based automation is also proposed earlier. Such sample-dependent automatic approaches may require setting some threshold values and underperform for abnormal data such as damaged anatomical structure etc., cases.

### 1.5. AI and process parameters

Several AI models are tested with the motivation to predict the intermediatory and/or initial protocols / hard controlling parameters by utilizing the process output in a feedback loop to drive the hardware[37], [38], [39]. A classical optimization approach minimizes the difference between predicted and desired outcomes (in case it is available) or simply minimizes an analytical model developed using previous data sets[40].

An AI model predicts the set of optimized process parameters based on its training data for the desired outcome. The accuracy to predict depends upon the diverse nature of the training data. The hardware control AI model generally depends on other AI models that are trained for feature extraction from the final data of best / desired characters. Sensitivity analysis for feature-extracting capabilities of available AI models has already been reported[41]. Most of the AI-assisted automation-related innovations in the field of OCT imaging happened for feature extractions to study the underlying variation (with respect to time and other parameters) in morphological and topographical structures using B-Scans [34], [41], [42], [43]. In general, for image segmentation U-net and variations are reported as all-rounder AI models [41].

A separate supervised Machine Learning model can be brought into the pipeline with a feature extraction AI model for optimal hardware alignment and process parameters' prediction. Feed Forward, Genetic Algorithms, Multilayer Perceptron, linear regression, feed-forward backpropagation, etc., are some of the preferred ANN models for process optimization. Particle Swarm Optimization, Artificial Bee Colony, Backtracking search algorithm, Gradient descent, etc., are the most used algorithm techniques [44]. Any study using 3D data (stack of B-Scans) from an OCT-SLO multi-spectral imaging system to train the ANN model for the same systems' process parameter optimization has not been reported.

## 1.6.    Motivation:

Quantitative functional analysis via imaging data, for example, estimating retinal/microvasculature thickness, requires accurate calibration. The sample specific self-calibration may be needed if media in the path of light or ROI composition changes during the measurement. In the case of an unhealthy retina, the optical properties along the axial path of light may vary, expecting us to estimate the axial resolution for every sample or after every scanning session. Enucleation for histopathological comparison is a worst-case scenario for clinical practice that is avoided for diagnostics and renders possibilities of time-dependent studies null. It may be possible that ROI composition and/or coordinates change with time. One example is imaging assisted photodynamic ocular therapy sessions. Varying ROI or beams' path may affect and change the calibration factors.

Accurate multi-spectral functional imaging is only possible if a system can automatically set its optimized operating parameters, including real-time hardware alignment, by receiving feedback via associated imaging metrics. The optimality of alignment is usually estimated using Enface structures or a single central slice of the ROI instead of total 3D volume. To assess if the alignment would give the best output in 3D, one should simultaneously estimate the finest structure in x-y and z directions. The finest structure that violates the corresponding spectrum's diffraction limit must be considered an error due to signal or image processing noise. Spatial resolution (axial and enface) thus must be estimated in real-time.

The novelty of this work is to test an OCT-SLO system design that is made fully automatic in such a way that process parameters can be optimized simultaneously to control multi-spectrum hardware. An adaptive design of an OCT reference arm integrated with a stepper motor and OCT-SLO beams focusing using a liquid lens controlled via AI algorithms pipeline to (a) analyze images and (b) set parameters are tested here to make the self-calibration process possible. Automatic variation of liquid lens voltage is also integrated with adjustment along the reference arm.

The comparison of relative accuracy between semi-automated and fully automated data acquisition requires stationary calibration phantoms/targets with already known dimensions. USAF target by Thor lab has already been used in several works. A novel target assembly is tested in this work for the coronal plane in the x and y directions. Transverse/axial plane spatial resolution is estimated using a novel method and a toy mouse.

Another gap filled in this manuscript is utilizing the swift automatic system alignment process to estimate sample independent calibration, axial, and Enface resolution.

In forthcoming sections, this work discusses a compact multimodal/multi-spectral system design and methods to sample dependent self-calibration method in the axial direction for accurate 3D functional imaging. In the end, retinal thickness is estimated.

## 2.  Materials and Methods

**Figure** 1(A) shows the imaging system's unique schematics and **Fig**. 1(B) shows its real-life photo with a mouse on stage for imaging position. Similar designs are reported earlier with minor differences with no description of stepper motor and liquid lens incorporations [45].

## 2.1. System hardware

### 2.1.1. Tabletop system case

The combined SLO and OCT imaging platform is custom-designed to take dynamic images of an adult mouse eye. The optics are aligned to cover the retinal section having a diameter of ~3.2 mm and a posterior nodal distance of ~2 mm with a field of view of $50^0$ degrees. The respective optics are fixed on breadboards/feature panels which are mounted inside a cage formed by rails with the help of support brackets. Rubber Anti-vibration dampeners are used between the joints. This cage fits inside an optically opaque enclosure of dimension 25 x 16 cm (shown in **Fig**. 1(B) top left). This box is made of Bead-Blasted Anodized Aluminium that is fit for darkroom experiments, and its breadboard is mountable (if needed) or can be fitted over a tiltable rack as well.

### 2.1.2. Sample mount

The optics are designed such that the scanning spot of the beam converges at 25 mm from the scan lens. Its location, however, can also be automatically varied by including a motorized animal platform or mount. A rotatable mouse stage with 5 degrees of freedom is designed to mount a bed to hold the anesthetized mouse in front of a horizontal scanning lens. A 3D printed mouse bed containing an aesthesia agent inlet and exhaust towards filters can be hooked on this stage. The double-sided ambidextrous cartridge/bed can easily be tilted and translated for precision ocular alignment with respect to the systems' optics. A single mouse imaging session includes (a) anesthetizing the mouse, (b) placing and stabilizing it on the mouse bed or mounting the sample on a scan lens mount, achieving optimal alignment of desired ROI/retinal location with system optics, and (d) saving data into PC for post-processing. It typically requires, on average, 30-45 minutes. The mouse bed is integrated with a heating pad to maintain the mouse's body temperature all the time; otherwise, its' cornea may suffer from the development of cold cataracts [36]. Cold cataracts may attenuate the light, thus affecting the quality of imaging. Please note that the mouse's upper body is also covered with a lightweight cloth or paper towel for long-duration imaging. A bite bar (also engraved in a 3D-printed bed) keeps the mouse's head stationary to provide its snout with the anesthesia agent. Gaseous isoflurane mixed with oxygen as a general anesthetic agent is preferred [36], [46], [47]. Its delivery tube is attached to an anesthesia vaporizing machine and an air pump. The system provides continuous and controllable delivery of an anesthesia agent to keep the mouse unconscious so that the eye remains available with just the right amount of ocular window to be probed by a Laser. Section 2.1.5 explains the mounting procedure for standard phantoms.

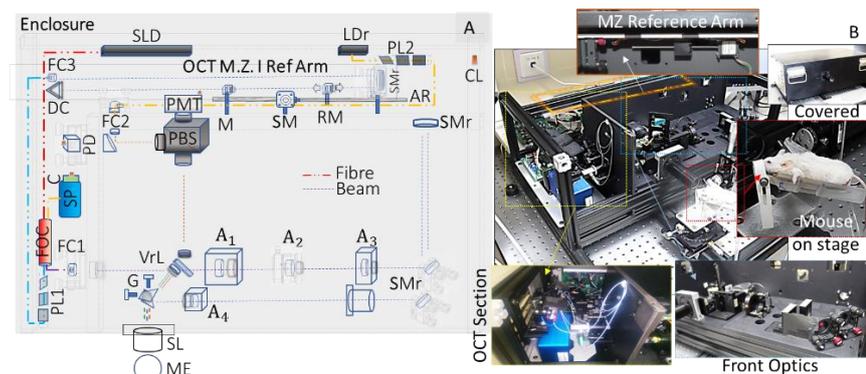

**Figure 1**: Multi-spectral Combined Imaging System; A) Schematic; Ax: Lens, AR: Actuator Rod, PBS: polarization Beam splitter & Fluorescence Filter Holder, CL: Camera Link Out, C: OCT Camera, D1: Detector, DC: Dispersion Compensation, DM: Dichroic Mirror, FC: Fiber Collimator, FOC: Fiber Coupler, G1&G2: Galvanometers, KM: Kinematic Mount, LDr: Laser Diode, its Driver & Thermal Control, RM: reference mirror, SL: Scan Lens, SP: Spectrometer, SM: Stepper Motor, SMr: Silver Mirror, PD: Photodetector, PMT: multichannel PMT, ME: Mouse Eye, VrL: Variable Focus Liquid Lens; B) Composite Photo of system with Mouse of imaging position.

### 2.1.3. Scanning Laser ophthalmoscope

As shown in **Fig**. 1(A), a 488 nm diode laser coupled with its drivers (LDTC 1020, Wavelength Electronics, Montana, USA) is used for confocal SLO reflectance and 520-550 nm fluorescence excitation. A combination of band-pass filters is mounted in the path of a Laser beam and coupled with PMT (Thorlabs Inc., NJ, USA PMM1001), explicitly measuring the green fluorescent protein (GFP) emission of 150 μW power along with reflectance in the visual spectrum. The interfacing codes allow to operate and switch between line excitations and raster scanning mode.

In the SLO part of the system, another reflective collimator (RC08FC-F01, Thorlabs Inc., NJ, USA) is used to launch the SLO light from a fiber with a polarization controller, such that the horizontally polarized light is reflected from a Polarization Beam Splitter (PBS, PBS251, Thorlabs Inc., NJ, USA). It is housed inside OEM Cube (TLV-UF-MF2 Thorlabs Inc., NJ, USA). The light was then reflected from a dichroic mirror (ZT405/488/561rpc-UF1, Chroma Technology Corp, VT, USA) to the silver mirror (PF1003P01, Thorlabs Inc., NJ, USA) and then co-aligned with the OCT light.

### 2.1.4. Optical coherence tomography

A fiber optic Michelson interferometer-based Fourier domain OCT is devised. The novelty aspect of the design is that the reference arm is coupled with a stepper motor, which the user can control by using the automation software interfaced with it, or it can set itself automatically without supervision. This feature is handy when a user is expected to align the reference arm non-manually, and optics is kept inside an enclosure to avoid contamination. The OCT light is split by a 90:10 850 nm 2x2 single-mode optical fiber coupler (TW850R2F2 by Thorlabs Inc., NJ, USA). The 90% portion of the light was connected to a reference arm consisting of a fiber collimator, a dispersion compensation block, and a mirror. The OCT probe beam was the 10% portion of light from the coupler, which was launched from a reflective collimator (RC04FC-P01, Thorlabs Inc., NJ, USA) and transmitted through a cold mirror (ZT670rdc-xxrxt, Chroma Technology Corp, VT, USA) for combination with the 488 nm SLO light.

The OCT beam uses compact Broadband Light Source Modules (superlum cBLMD) with a 50nm bandwidth centered at 840 nm and delivers ~643 μW at the mouse pupil plane, good enough to acquire ~2μm theoretical axial resolution in tissue. A spectrometer (Cobra 800, Wasatch Photonics Inc.) with a high-speed line CMOS camera (Octoplus Teledyne e2v) is used as the OCT detector.

The Variable Focus Lens (variable focus liquid lens, varioptics, Corning, NY, USA) creates the entrance pupil plane of both OCT and SLO sections. The imaging beams are relayed to a mounted pair of Galvanometer-scanning Mirrors (GM, 16-039-dual-axis Saturn, Edmund optics, MA, USA) with a clear aperture of 3.0 mm. A beam diameter of 1.0 mm is relayed from the GM towards the scanning lens (SL, WPQ10E, Thorlabs Inc., NJ, USA) and on the mouse retina at its end.

The backscattered OCT light from the mouse eye is recombined with light from the reference arm via a fiber optical coupler and directed to a spectrometer. Five hundred line/A scans are acquired using an OCT Camera mounted directly on the spectrometer. The OCT volume is sampled into 2048x500x360 dimensions. Multiple B-Scans (either 3 or 6) are saved for phase variation detection (due to fluid flow) and to develop optical coherence tomography angiography (OCT-A). DAQ Board (PCIe – 6361, NI Texas, USA) is used to simultaneously transfer the SLO and OCT Data into PC Card (Xtium-CLMX4, Teledyne Dalsa, Ontario, Canada).

### 2.1.5. Phantom for calibration

A positive pattern USAF target (a standard calibration phantom) is used to devise an Enface calibration Phantom. Its optomechanical, including individual components, is shown in **Fig**. 2(A). Lens Tube (SM1L15) houses the resolution test target (1951-R1DS1P) with the help of a brass spacer (SM1S10M) in front of the aspherical lens (C105TMDA). The target is packed from another set of spacers (SM1S01 and CMRR) from behind with the help of an end plug (SM1PL). The mentioned components are procured by Thorlabs. The inside surface of the end plug is pasted with a fluorescence compound (FMG - Green Fluorescent Polymer Microspheres 1.3g/cc - 1-5μm, Cospheric LLC, CA, USA) that helps to test emission in the form of a target pattern [49], [50]. The negative equivalent pattern etched on the glass surface of the target is shown in **Fig**. 2(B). A standard resolution chart (shown in **Fig**. 2(C)) is used to find the correlation between the length of the strip seen in the image and its real length etched on a pattern. Values of 'x' are specified in the manufacturer's data sheet [48].

## 2.2. Image Acquisition

The system is built to acquire multiple B-Scans belonging to the same location. This, although is a memory intensive time taking process. It, however, facilitates statistically rich information that is used to create

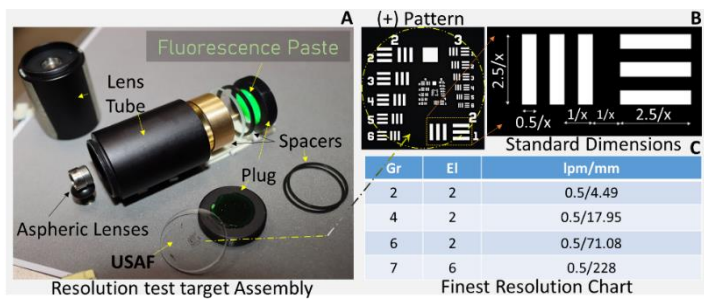

**Figure 2**: Enface calibration Phantom: (A) shows components including USAF; (B) highlights the True Design of the Target and dimensions; it also shows an approximate field of view available to the scanning lens of the system by the yellow dotted circle; (C) shows estimated spatial resolution for all three spectrums, finest is 2 microns.

OCT-A images later (if needed). In total, 1080 B-Scans are saved for particular sections of the ROI/eye and averaged into 360 B-Scans for improvement in Signal to Noise Ratio (SNR) by reducing the scattering inhomogeneities [49]. The respective optimal and sub-optimal alignment settings (for each case) are iteratively varied using a manual approach and recorded as well. The intermediatory images (including best) are recorded for three different samples: (a) mice, (b) USAF assembly, and (c) toy mouse. The detail of the samples is given in the result section.

An AI/SNR optimization-based image processing code is integrated with a stepper motor and voltage of variable lens controlling codes. The control code varies the stepper motor position and voltage of the lens in the first step. Simultaneously, intermediate data: (a) converted RAW into 3D volume/360 B-Scans, (b) respective arms position, and (c) voltage of lens is acquired in real-time. Acquired 3D volume can be displayed as an option. Afterward, another part of the image post-processing code compares the 3D volume data set for all optical alignments. The comparison stage has two options (to select a priori) either: (a) a pre-trained machine learning model or (b) a maximum signal-to-noise analysis-based process. Two options are possible to quantify the best parameters/alignment: (a) based on 3D volume by processing all 360 B-Scans at once or (b) simply processing averaged (over 3D) Enface image. The latter may discard the significant information in the axial direction, resulting in inferior results. It, however, requires low processing power and would be relatively faster. The automatic controlling code can be decoupled to provide manual override. In this work, the former option is thus chosen. Once the optimal setting is obtained, axial and enface calibration is performed for the sample automatically, and the final data set is saved along with its corresponding optimal settings. The process flow is shown in **Fig**. 3.

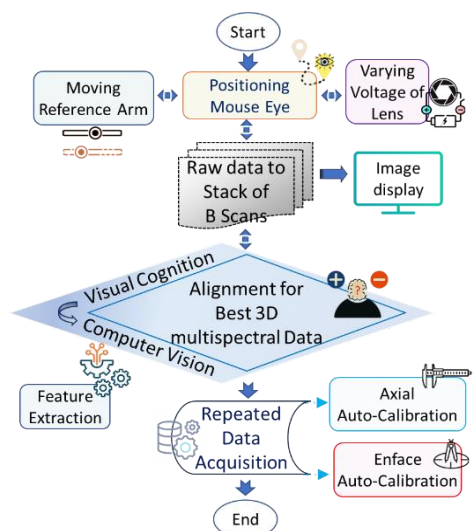

**Figure 3**. Process Flow for self-calibration and automatic imaging.

## 2.3. SNR-based hardware parameter Optimization:

A classical method estimates the averaged value of Signal to Noise (SNR) of 3D volume from all alignment settings and selects the corresponding hardware settings that give maximum SNR. The SNR in decibels relative to the carrier (dBc) of a real-valued sinusoidal input signal is determined using a modified periodogram of the same length as the input[50]. Kaiser–Bessel window trade-off parameter between main-lobe width and side lobe level ∝ value 12.1 is used[51]. The power of the first six harmonics, including the fundamental, is excluded.

## 2.4. AI-based Hardware Parameter Optimization:

Two options are possible to integrate AI with hardware control codes to predict operating/process parameters affecting associated hardware alignment/configuration: (a) either scan between extremums, let AI process image index-based parameters, or (b) let AI receive a-priori information about the sample from the operator and let it make a straightforward decision. Since the latter option may not work for all samples, the former is tested in this work.

For the image index-based parameter process two options: choosing a 3D volume with a maximum (a) number of features or (b) similarity index is possible. The latter approach uses semantic segmentation and requires a pre-trained apt data set. Moreover, to estimate the number of features, one may use the entire B-Scans stack (3D volume) or simply average them to use Enface. For SNR driven automation, both are used, but for AI, full-stack is used.

AI models, such as U-Net, ResNET, etc, can be used for similarity index estimation. The sensitivity analysis of AI models is not the primary goal of this work; thus, detail is skipped here for brevity. Readers can find in-depth analysis elsewhere[41]. U-Net is trained (using 20 Balb females' mice retinal data) to identify the best 3D volume consisting of the most distinct structure of mouse retinal layers [52], giving 98% accuracy but found to be relatively slow for real-time use. Its performance significantly depends on testing and training data, and it may not work if abnormal features are present in ROI. The PC's CPU attached to the multi-spectral imaging system is expected to control several hardware drivers, acquire the RAW data, and process its 3D volume sequentially. Integrating another sequential processing step to estimate the similarity index may further slow the overall performance.

The former method estimates total features, irrespective of the nature of those features in 3D volume, thus also tested in this work. MobileNet v2, another AI model optimized for low-latency, low-power processing units, is tested [53]. A study shows good agreement between human observers, model observers, U-Net, and MobileNet[54]. MobileNet extracts the features from any given image and returns a large set of different features for each image. The extraction process of each image remains the same. Simply total sum of these features is used as a scoring mechanism. MobileNet is loaded from TensorFlow. A dedicated function, "analyse_image" is written (attached to this article). This function first normalizes the pixels of all the 360 B-Scans, then converts them to an array, adding an extra dimension to the B-Scans set. Each B-Scan is pre-processed using the MobileNet model to obtain a feature that acts as a distinguisher. The "find_best_image" function takes a list of image paths as input and iterates over each B-Scan path to find the B-Scan with the highest score. The B-Scan set with the highest score (sum of features) obtained from the MobileNetV2 model is chosen as the best image set. The code iterates over a list of image paths and analyses each B-Scan using the analyze_image function. The analyze_image function loads the B-Scan, pre-processes it, and passes it through the MobileNetV2 model to obtain a feature vector. The sum of the feature vector is then calculated using np.sum(features). The hardware interfacing code in sequence keeps varying the lens's reference arm position and voltage (from one extremum to another) while acquiring and saving the successive sets of B-Scans. The mentioned AI model keeps estimating respective feature vectors for each alignment. The main code keeps track of the highest score encountered so far (best_score) and the corresponding B-Scans set path (best_image). If a higher score is found for a subsequent B-Scan set, the best_score and best_image variables are updated. In the final step, the stepper motor moves the arm to the position where this best B-Scan image is acquired, and the corresponding voltage of the lens is set. The scanning process is shown in M1 or Movie 1. AI pipeline is used for retinal layer thickness estimation [52].

## 2.5. Calibration of SLO-OCT system

Typically, a system is calibrated to estimate how fine a structure it can probe. The calibration process requires a phantom with physical structures of several orders of known dimensions and their corresponding number of pixels in its image obtained using that system. The calibration factor, a ratio of the dimension of the finest structure and the number of pixels in its corresponding image, provides system resolution. The pixel-to-millimeter calibration provides a means to estimate the dimensions of structures in an unknown sample. The only constraint is that the path's optical properties till ROI in both cases (known Phantom and unknown sample) must remain the same. The calibration factor otherwise would be sample dependent. It is an impractical requirement and thus has limited utility. The calibration factor for the same Phantom may also differ for axial and enface direction. The following subsection describes the process of estimating both factors.

### 2.5.1. In the x-y / En face direction

One simple method utilizes the sample thickness itself. Suppose the sample's thickness is known (for example, balb/c mouse retina is ~2 mm), then the voltage range can be correlated with the thickness of the sample, which can further be used to estimate the axial resolution. This method does not guarantee to resolve the finest structure in ROI. Change in lens voltage and the displacement of illuminated pixels thus can be used as an alternative to estimate the axial resolution of the sample. Variation of lens voltage, its relationship with width of the sample (provided it is known, for example, the approximate thickness of any retinal layer, etc.,) and correlation factor / axial resolution estimation can be simply automated and can be made sample-independent using this method.

### 2.5.2. In the Z / axial direction

For z-axis resolution estimation, using a biological sample such as a retinal image may provide less reliable physical dimensions when compared with the real world. For example, histopathology data is less accurate. For regular and abnormal retinas, the composition of the path may significantly vary in axial direction. A sample for which the axial direction physical dimensions are not even known is useless. *In this work, the systems' reference arm measurements itself is utilized to estimate the sample-specific calibration factor.* When the reference mirror is moved by rotating the threaded arm, features in the acquired image are observed to move away. If one can quantify the distance traveled by the mirror and the number of pixels with which a certain feature in ROI is displaced, one can quantify the Z-axis spatial resolution. Acquiring the best data set (having the finest axial direction and the finest x-y direction) is done in real time while varying all involved alignment parameters using 3D volume. Moving the reference arm only affects the OCT spectrum; thus, this method provides axial resolution in the OCT spectrum only.

### 2.6. Curvature control

An artificial curvature (in OCT B-Scan) in the axial direction may exist in the structure under ROI for a suboptimal alignment. An example shown in **Fig.** 4 illustrates this point. Two images of the mouse retina are captured using OCT when two different combinations of reference mirror position (RM) and distance of the eye from the scanning lens (SL) are set one by one. The distance of the mouse eye from the scanning lens (SL) is measured using a micrometer that translates the mouse platform (a similar description is given in section 3.2 and shown in **Fig.** 6). The reference arm position and thus the translation of the mirror mounted on reference arm (RM) is also measured using the number of turns that are calibrated on actuator rod into the linear distance. These images are shown in **Figs.** 4(A) – 4(D).

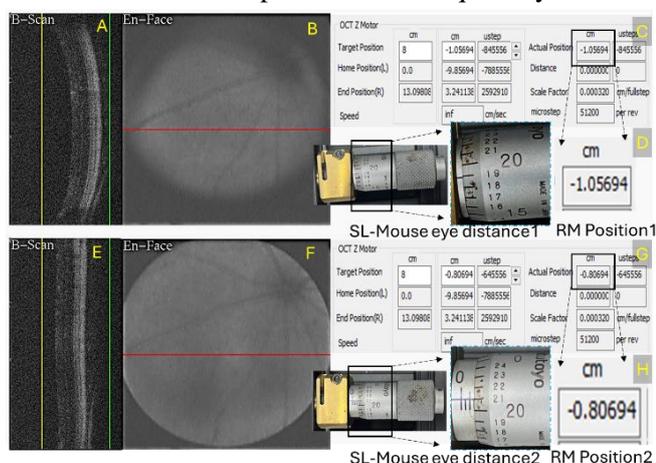

**Figure 4.** Effect of relative position of mirror on reference arm (RM) and distance (ROI-SL) of mouse eye (ROI) from scanning lens (SL): (A) and (B) shows averaged Enface and B-Scan *with curvature* corresponding to horizontal red line, (C) and (D) shows corresponding actual position of RM and distance between ROI and SL; similarly, (E) and (F) shows improved (but not the best) Enface and flat B-Scan, when (G) positions of RM and (H) ROI-SL is changed.

Suppose one moves the mirror on the reference arm and mouse eye from the scanning lens for a certain combination. In that case, the B-Scan (section highlighted using a red line on the enface) becomes relatively flat, and the quality of the Enface improves. It is depicted in **Figs**. 4(E)-4(H), respectively. It shows that the curvature in B-Scan can be flattened out (if needed) by adjusting the relative position of the reference arm mirror and axial distance of ROI along the laser from the scanning lens, mutually. It is quite possible that the retina may not have curvature as depicted in **Fig**. 4(A) for the small size of ROI. It is also depicted that the B-Scan shown in **Fig**. 4(E) is not the flattest as at both ends (top and bottom) it shows slight concavity; thus, a shadow appears in En-face at corresponding spots.

*The best scan quality in the region of interest (inside the eye) is, thus, achieved by visually analyzing the image quality in real-time while adjusting the: (a) voltage of the liquid lens (if used), (b) length of reference arm and/or (c) axial distance of ROI/mouse eye along the laser beam, simultaneously.*

## 3. Results

Three different samples are used: an enface target phantom (as it does not have variation in the z-axis), a toy mouse with a fluorescent glass bead in place of its eye, and balb/c mice. The first step describes the classical SNR-based automation process for Enface calibration. Afterward manual process to generate the training data set for AI control codes is explained. The data is used to train AI, and its performance is tested next. The automatic calibration steps described in section 2.5 are tested with an AI-driven mechanism. The mice imaging results are presented at the end.

### 3.1. Calibration

A typical method to characterize an in-vivo imaging system is to calibrate its images with images taken by an analogous ex-vivo imaging system. Spatial resolution is the correlation between these two techniques. It is obtained just by taking the ratio of the dimension of a particular structure in meters with the number of pixels used to depict that many dimensions. The calibration process expects that the imaging system is aligned perfectly. Thus, the method explained in section 2.5 is tested first. Only OCT spectrum results are used to show the performance for the sake of clarity and brevity.

### 3.1.1. Coronal/frontal/ lateral/enface/x-y plane spatial resolution

As described in section 2.5.1, the USAF assembly (shown in **Fig**. 2) is mounted on SM, and the voltage of the liquid lens is varied. **Figure** 5 depicts the effective shift in peaks in the sum of A-Scans, quality of B-Scans, and Enface of OCT spectrum corresponding to variations in lens voltage as shown in **Figs**. 5(A-D), respectively. **Fig**. 5(E) shows the binarized line structure depicting that variation in lens voltage (values are specified in each subfigure as well) loses distinctions between horizontal lines, and an optimal lens

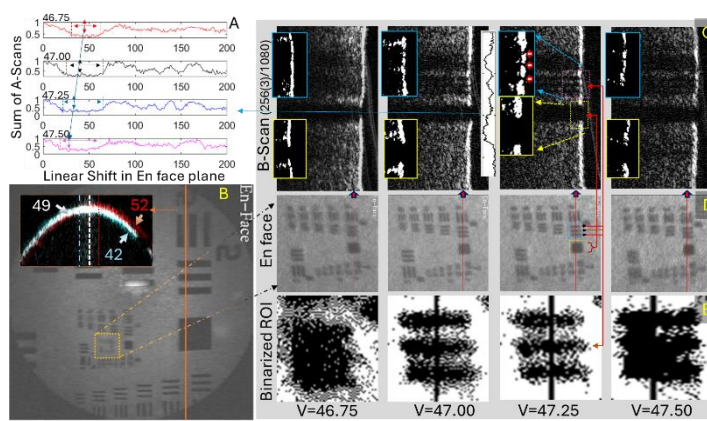

**Figure 5:** Effect of variation in voltage of liquid lens, (A). Sum of A-Scans along axial direction of first zoomed ROI shown in (B). show sharp peaks with precise width in A-Scan correspond to voltage 47.25, (B). also shows shit in B-Scans corresponding to vertical orange line in inset when lens voltage is changed from 42, 49 and 52 volts in Enface direction, (C) shows B-Scans of zoom first ROI corresponding to red vertical line in enface images shown in (D), (E) shows binarized zoom ROI inside first zoom ROI. The three horizontal lines gets blurred into each other for 46.75, and 47.50 volts. Showing that lens voltage shifts the focus in En-face plane; 47.25 third row in (E) is optimal settings. Experiment recording is M2 https://youtu.be/f1guvHYOd-4

voltage (47.25 units) captures the best features. The process of variation is captured in an M2.

### 3.1.2.   Estimation of Axial/transverse/Z Plane spatial resolution:

The depth of the structures on the USAF target surface is not accurately known, so it cannot be used for axial plane spatial resolution estimation. Estimation of axial spatial resolution of OCT imaging system is reported using full retinal thickness (between internal limiting membrane till the end of the Bruch's membrane) is 2.09 mm[33], [55]. However, as the light travels, the focal spot size/ point-spread function size varies in tissue due to attenuation. So, these methods only provide axial resolution in retinal tissue. The utility of the OCT imaging technique is also proven in estimating cell population density and size in vitreous humor. In most cases, moving the mouse mount or sample away may not be a good idea once the liquid lens's reference arm position and voltage values are optimized. Instead, as proposed in section 2.5.2., the actuator rod is rotated five rounds in both directions and physical displacement in terms of displacement of pixels is noted. Since the distance traveled by the reference mirror in the actuator rod is already fixed and known, the ratio of this value with the number of pixels is estimated. **Figure** 6 explains an experiment to estimate axial resolution using a toy mouse eye and a fluorescence glass bead. It is shown in **Fig**.6(A) top left inset position. Its' eye is made of glass having homogeneous material distribution; thus, insignificant variations in refractive index are expected to have a single calibration factor in the axial direction. This toy mouse is mounted on a platform with 5 degrees of freedom (can be varied using integrated micrometers attached to micro positioners in respective directions), including all three axes for desired alignment. The optimally aligned position (as baseline (Z = 0 mm)) is shown in **Fig**. 6(A). Its corresponding micrometer position and B-Scan are shown in **Figs**. 6(D) and 6(E). In the second and third stages of the measurements, the platform is moved 1.27mm (0.05") and 2.54mm (0.1") away from the baseline without disturbing alignment in any other direction. Corresponding micrometer positions and B-Scans (center section highlighted using blue both side arrow line on enface) are shown in **Figs**. 6(C), 6(B) and 6(E).

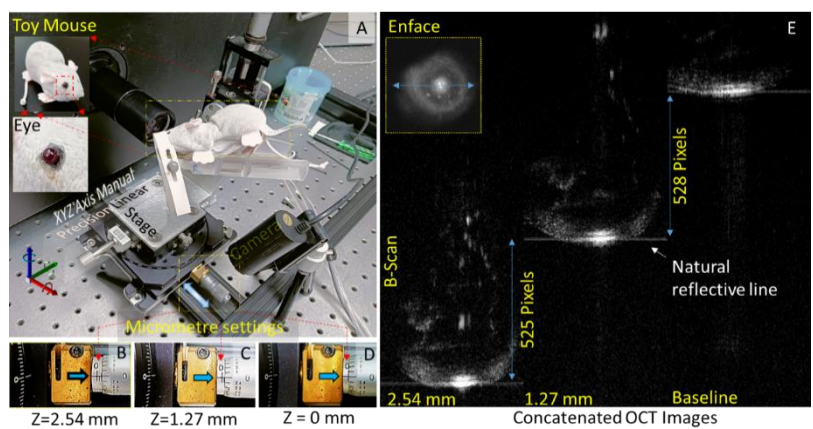

**Figure 6**: Calibration in the axial plane via OCT B-Scan; (A) shows a toy mouse with a glass bead eye being scanned and part of the frontal view of the multi-spectral animal imaging system, (B-C) micrometer head positions shows the extent of displacement of a focused eye away from the reference position in (D), (E) shows corresponding OCT images and enface. Experiment recording M3: https://youtu.be/6QOLbq9ajyQ.

The experiment is recorded in the form of supplementary M3. All three B-Scans are concatenated side-by-side to visualize the change in position. The eye B scan shows a naturally existing reflective horizontal line. It is used as a reference to estimate the equivalent number of pixels in B-Scan images corresponding to the physical movement of the eye. This experiment is repeated eight times, and just one instance is shown. The average number of pixels corresponding to each 1.27mm movement is 526, giving us an axial resolution equal to 2.41 microns in a glass bead synthetic eye. Lens voltage is kept constant during this experiment.

Results shown in section 3.1 confirm the automatic methodology to find sample-dependent 3D spatial resolution thus the finest structure that can be used to calibrate the system. In the next section, the automation procedure is tested.

### 3.2.   SNR-based optimal setting:

After testing the calibration methods manually and successfully, this section discusses automation methodology. The analysis utilizes steps explained in section 2.3, and results are shown in **Fig**. 7. The Enface target phantom assembly (shown in **Fig**. 2) is mounted on the scan lens of the imaging system. Full ROI is selected. The real-time control for the M-Z arm, liquid lens voltage, and corresponding B-Scan (center slice of 3D volume) are captured in the form of M4 and M5, respectively. The averaged SNR value is used when a full B-scans stack is considered. Otherwise, the SNR value is considered for a single Enface. The standard deviation when the average value is used is not insignificant. It is, however, not shown on the same plot so that characteristics of the SNR plot can be observed. The top row of **Fig**. 7(a) shows a final instance of scanning results. B-Scan (BSn$_{RealTime}$), Enface (EnF$_{RealTime}$), and SNR values (SNR-BSn and SNR-EnF) are shown when the M-Z arm is moved between extremums. Lens voltage is kept constant at its center value. The maximum value of SNR is highlighted using a circular-shaped marker with a dot at the center of the same plot. On the bottom row, the first two images show B-Scan (Best-BSn$_{Bs}$) having maximum SNR when SNR is calculated using B-Scans stack, its corresponding Enface images (Best-EnF$_{Bs}$) till that moment. The last two images in the bottom row show B-Scan (Best-BSn$_{EnF}$) and its parent Enface images (Best-EnF$_{Bs}$) corresponding to the maximum Enface when SNR is calculated using Enface only. Both approaches when: (a) Enface image or (b) full B-Scan stack is used to estimate maximum SNR generate inferior performance when variation in image features are relatively large. The relatively large variation (from only background speckle noise to structure related to USAF target and, again, speckle noise only) in images depends on the thread count in the actuator rod (AR). The stepper motor/actuator rod rotated 19 times between -4 to -2 cm.

A similar analysis is performed, but this time varying lens voltage and keeping arm length constant. Its equivalent results are shown in **Fig**. 7(b). The presence of identifiable signatures of USAF target in B-Scan and Enface acquired in real-time images (top two images) show less variation. Irrespective of either using B-Scan stack or Enface image for estimating maximum SNR and its corresponding hardware setting results in relatively better images as shown in bottom row images. Maximum SNR (-1.98 dBc) is obtained by using an Enface for SNR estimation. It shows SNR, in particular, is useful when variation in lens voltage causes very small changes in image features. This is possible when ROI has fewer features in the axial direction.

The SNR (when the B-Scan stack is used to estimate maximum SNR) varies between -18 to -10. The SNR (when Enface is used to estimate maximum SNR value) ranges between -20 to 0 with a standard

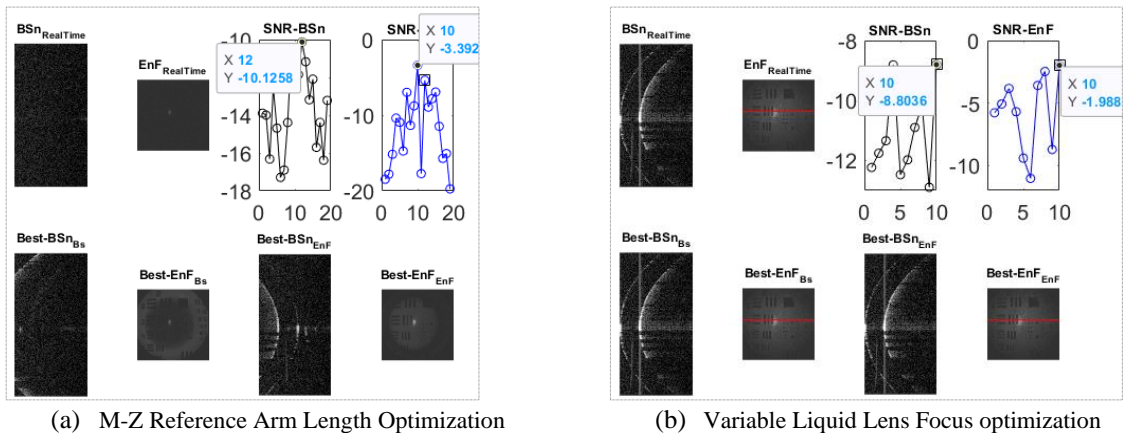

(a)  M-Z Reference Arm Length Optimization  (b)  Variable Liquid Lens Focus optimization

**Figure** 7: Imaging Parameters Optimization using SNR in Real-Time. M4: https://youtu.be/b6pwhDN1b9M & M5: https://youtu.be/S1Cib-5M2k0 illustrate real-time output. Both approaches generate inferior results.

variation of 2.15 and 4.9. This data corresponds to variation in the M-Z arm while keeping lens voltage constant. This variation (B-Scan: -13 to -8 and Enface: -11 to 0) with a standard deviation of 1.4 and 3.0, respectively, is smaller if lens voltage is varied. It shows that an Enface image (an averaged coronal image of 360 axial B-Scans) as a reference to estimate maximum SNR offers more sensitivity relative to a full stack B-Scan.

### 3.3. AI acquired optimal setting:

#### 3.3.1. Training dataset generation

The USAF assembly is mounted on the scanning lens. A manual procedure is followed to find the optimal settings of the length of the reference arm, voltages of the liquid lens, and PMTs. The reference mirror (RM) is slid on the actuator rod (AR) by hand from one end to the other while keeping the. PMTs are kept on maximum value. The voltages of variable lenses are also kept fixed on 45 units. Real-time B-Scans giving the impression of the cross-section of the USAF is anticipated on the display screen. Once a structure appears, it is brought to the center of the screen by further adjustment. The image (composite of B-Scan and Enface averaged between green and yellow vertical lines on B scan) shown in **Figs**. 8(A) to 8(E) depict various stages. The corresponding fluorescence images are also shown next to the Enface image in the same composite image set. The inset image also shows the actual position of the reference arm on the actuator rod. The USAF B-Scan is missing in **Fig**. 8(A) and starts appearing when it is moved from -3.79 to -3.34 cm. The best position appears at -3.26 cm, shown in **Fig**. 8(c). It is shown in **Figs**. 8(D) and 8(E), that if the arm is moved further in the same direction, the lens disappears again, giving a blank Enface. **Figures** 8(F) to 8(i) show B-Scans, corresponding Enfaces, and fluorescence images appearing when lens voltage settings are varied between 46.38 and 47.00 units. RM position of **Fig**. 8(c) is used. It seems with all these variations best image acquired is in **Fig**. 8(i). The reference arm setting is used as depicted in **Fig**. 8(c). The RM at -3.26 cm with voltages at 46.38 and 47.18 units may be accepted as the final optimal settings. More than 100 such intermediary images are acquired; however, only a few are shown here. These intermediary images are now used to train the AI model to identify the sharpest B Scan.

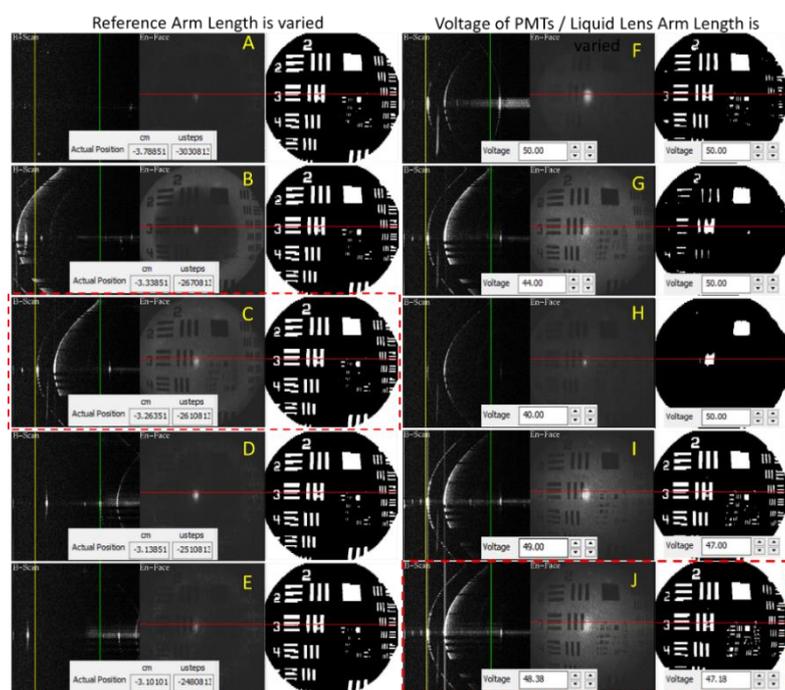

**Figure** 8: Generation of training data. A-E: reference arm length is varied manually, F-J: voltage is varied keeping arm setting of 8(C).

#### 3.3.2. AI driven imaging

This model is integrated with stepper motor-variable lens control and data acquisition codes. The code, as expected, acquires possible extreme locations referred to as initial (P1 position indicated by yellow arrow) and end-stage (P2) in **Figs**. 9(a) and 9(d). Their corresponding B-scan, enface images, and reference arm positions are shown in **Figs**. 9(b), 9(c) and 9(e) and 9(f), respectively. The B-Scan and enface are all blank. The AI mode is selected so the motor has slid through between both ends while saving and analysing the intermediatory images. Based on the previously trained model, it figures out the best-acquired image and realigns the reference arm (on P3 position) and PMTs on corresponding settings of -3.3 cm and 47.01 and 47.09 a unit. If we compare images acquired manually and using the AI-driven method, the image shown in **Fig**. 9(J) is inferior to the image shown in **Fig**. 9(h) in terms of sharpness, reflection, and clarity. The whole process is recorded and presented in the form of [M6](#) (for USAF) and [M7](#) (for Toy Mouse). This AI-driven, fully automated approach is used to calibrate this system in the next stage. The U-Net performance is found to be training data dependent and has a significant computational cost; thus, MobileNet is used. The best image returned, in our case, 1280 features. The displacement of the actuator rod in the case of AI control is less than 1 cm, which is relatively smaller when it is moved using SNR-controlled code. Similar settings, when acquired by SNR-based approach arm moved 2 cm between -4 to -2 cm. Assuming traversing between extremum is the same, the AI method is 65% faster. To reach these settings, it took more than 10 minutes. The arm moved 0.7 cm from -3.8 to -3.1 cm in less than a minute. The SNR of AI detected best B Scan and Enface is -8.8 dBc and -5.7 dBc.

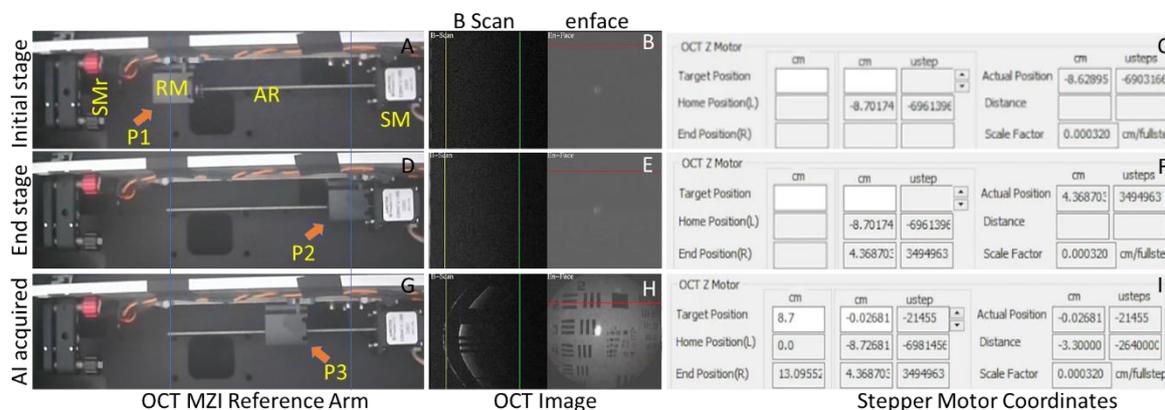

**Figure 9**: Automated OCT setting by AI; RM: reference mirror, SM: stepper motor, AR: actuating rod, P: location of RM; (A) reference mirror on its initial position in the extreme left, (B) B scan (corresponding to the red horizontal line on enface) and enface (averaged between green and yellow vertical lines shown on corresponding B Scan), (C) coordinates of stepper motor, (D) reference mirror on its end position near to stepper motor, (E) and (F) corresponding B Scan, enface and stepper motor coordinates, (G) reference mirror on AI optimized position, (H and I) corresponding B Scan, enface and stepper motor coordinates. Experiment recording M6: https://youtu.be/rJkWEJ8hM10 (USAF) and M7: https://youtu.be/99s2t8JjkM4 (Toy Mouse).

## 3.4. Automatic Imaging

Images from all three spectra have the same lp/mm spatial resolution correlation. The dark black (chosen fake color) region would be due to fluorescence paste emitting, and the white part is due to the pattern blocking it. Transmitting direct and OCT images would be the same as the target light color due to glass and dark due to pattern. The Phantom is mounted on the scanning lens of the system. Figures 10(A-L) show images of the multi-spectral/combined system acquired simultaneously. **Figure 10** shows images acquired using AI control and processing codes. The real length of the marked (using a thin yellow arrow) stripe is mentioned at the top of each fluorescence image. Images shown in **Figs**. 10(A-D), **Figs**. 10(E-H) and **Figs**. 10(i-L) are acquired by selecting different zoom levels for the respective spectrum. A rectangular ROI is drawn on **Figs**.

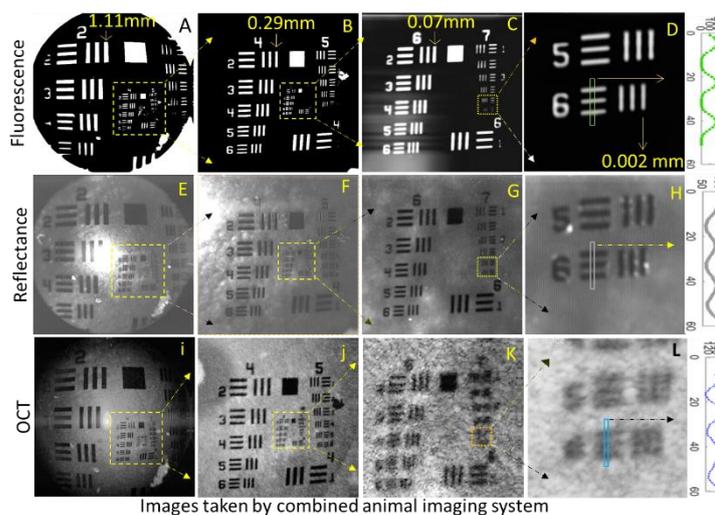

**Figure 10**: Spatial Resolution estimation of multi-spectral Imaging System: (A-B) shows 450nm Fluorescence Layer reflection acquired, each image contains the width of line pointed by an arrow; (E-H) shows reflectance images acquired; (I-L) are images acquired from 890nm OCT Enface images; (M) shows line plot of the vertical rectangular region (5 pixels wide and 60 pixels long) on 6th Element (El) of 7th Group (Gr), respective images (D, H, and L) show the region of interest in green, grey and blue colour rectangle marker. The finest structure of 2 micron is identified.

10(D), 10(H), and 10(L), and line plots of each spectrum are plotted in **Fig**. 6(Q), showing that fluorescence and reflectance images have uniform pixel distribution recovery capability. It is shown that this system can measure 2 microns (the finest stripe pattern etched on USAF target) with clarity. OCT image has little distortion at this level of zoom, but the pattern is visible with clarity. We note that the OCT image enface is a result of averaging 359 B-Scans. Structures finer than axial spatial resolution will not be averaged and thus fail to appear in enface. Thus, axial spatial resolution may affect frontal plane spatial resolution, which inspires and highlights its serious assessment.

### 3.5.    Application on animal ocular imaging

After building sufficient confidence towards AI driven automation, the technology is applied for mouse ocular imaging. Several animal disease models require a time-dependent variation of retinal thickness as one of the imaging biomarkers; thus, retinal thickness measurement is targeted as the final result of this work. The institute's animal ethical committee has approved the protocol for animal housing under protocols BT/IAEC/2018/2/03 and BT/IAEC/2023/2. Twenty mice (Balb/c female mice (average age of 1 year)) are used to estimate the retinal thickness without induction of any disease and kept in pristine environments.

Abnormal structures may appear if the animal is suffering from a disease. This may or may not be present in training data. MobileNet is replaced in the U-Net model in the AI pipeline for post-processing and retinal segmentation[34]. Following the steps described in section 3.2.2 above, the 1.27 mm translation created a displacement of ILM equal to 492 pixels, giving an axial resolution of 2.58 microns in vitreous humor. **Figure** 11 shows the images. **Figures** 11(a) and 11(b) shows the OCT and OCT-A enface retinal images of one of the mice. The yellow-dotted rectangle with a blue center line with an arrow in **Fig**. 11(a) shows the B Scan in **Fig**. 11(c). **Figure** 11(d) shows the number of pixels counted using AI (verified manually with negligible error) throughout the layers. Similarly, 2.1 mm translation makes the retina disappear from the screen, thus giving an axial resolution of 8.4 microns as the front of the retinal layer to the back of the sclera is made of 251 pixels. We note that the amount of translation required to make the retinal image totally disappear from each B Scan varies and requires significant efforts, making this approach less practical to implement despite being accurate. The average pixel number (including every B Scan of twenty mice) is $275 \pm 45$ pixels; thus, the average axial resolution is $0.76 \pm 0.46$ microns.

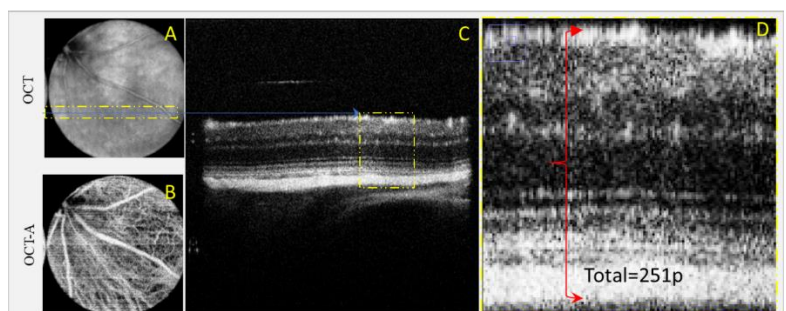

**Figure** 11: Axial resolution estimation of imaging system for mouse eye.

## 4.  Conclusions

AI-automated multi-spectral ocular imaging system design is presented here. The stepper motor integrated Mach Zehnder reference arm and voltage driver of a liquid lens is integrated with an image processing AI code in a negative control loop. The image processing AI code can identify desirable features in the image while the arm moves, and lens curvature is varied to focus the beam with minimum dispersion and best contrast. This intelligent automation takes place in real-time and gives accurate hardware settings faster than that can be achieved manually otherwise. The slow alignment might result in discomfort to the patient or create a situation of cold cataracts. It is shown that an alternative classical approach that optimizes the signal-to-noise ratio of acquired image works only with lens control when feature variation is relatively small. The limitation of AI model sensitivity to identify new abnormal features in disease-ridden retinas can be overcome using a robust model trained on data with several features. Self-calibration procedures in the x, y, and z-axis steps are shown to be working successfully after the optimal alignment procedure.

## Declaration

*Institutional Review Board Statement:* The animal imaging sessions are performed under approved protocols BT/IAEC/2018/2/03 and BT/IAEC/2023/2 compliant with the CPCSEA, Govt of India, and the approval of the Institute Animal Ethical Committee at IIT Roorkee.

*Availability of data and materials:* The code for image AI optimization and Data will be provided on request.

*Funding Acknowledgments:* IITR/SRIC/1073/IMPRINT-2 and DST-SERB: IMPRINT-2, IMP/2018/001045.

*Competing interests:* Indian Patent office allotted application no. **202411011899**.

## Reference

[1]  A. F. Fercher, "Optical coherence tomography – development, principles, applications," *Z Med Phys*, vol. 20, no. 4, pp. 251–276, Nov. 2010, doi: 10.1016/J.ZEMEDI.2009.11.002.

[2]  R. H. Kardon, "Role of the macular optical coherence tomography scan in neuro-ophthalmology," *Journal of Neuro-Ophthalmology*, vol. 31, no. 4, pp. 353–361, Dec. 2011, doi: 10.1097/WNO.0B013E318238B9CB.

[3]  J. Welzel, "Optical coherence tomography in dermatology: a review," *Skin Research and Technology*, vol. 7, no. 1, pp. 1–9, Feb. 2001, doi: 10.1034/J.1600-0846.2001.007001001.X.

[4]  "The Role of Optical Coherence Tomography in Coronary Intervention FAU - Terashima, Mitsuyasu FAU - Kaneda, Hideaki FAU - Suzuki, Takahiko," *Korean J Intern Med*, vol. 27, no. 1, pp. 1–12, Mar. 2012, doi: 10.3904/kjim.2012.27.1.1.

[5]  S. Barry *et al.*, "Comparison of three-dimensional optical coherence tomography and high resolution photography for art conservation studies," *Optics Express, Vol. 15, Issue 24, pp. 15972-15986*, vol. 15, no. 24, pp. 15972–15986, Nov. 2007, doi: 10.1364/OE.15.015972.

[6]  R. D. Ferguson *et al.*, "Multimodal adaptive optics retinal imager: design and performance," *JOSA A, Vol. 29, Issue 12, pp. 2598-2607*, vol. 29, no. 12, pp. 2598–2607, Dec. 2012, doi: 10.1364/JOSAA.29.002598.

[7]  N. Kraus, F. Placzek, and B. Metscher, "OCT Meets micro-CT: A Subject-Specific Correlative Multimodal Imaging Workflow for Early Chick Heart Development Modeling," *Journal of Cardiovascular Development and Disease 2022, Vol. 9, Page 379*, vol. 9, no. 11, p. 379, Nov. 2022, doi: 10.3390/JCDD9110379.

[8]  N. Kraus, F. Placzek, and B. Metscher, "OCT Meets micro-CT: A Subject-Specific Correlative Multimodal Imaging Workflow for Early Chick Heart Development Modeling," *Journal of Cardiovascular Development and Disease 2022, Vol. 9, Page 379*, vol. 9, no. 11, p. 379, Nov. 2022, doi: 10.3390/JCDD9110379.

[9]  R. J. Zawadzki, A. R. Fuller, D. F. Wiley, B. Hamann, S. S. Choi, and J. S. Werner, "Adaptation of a support vector machine algorithm for segmentation and visualization of retinal structures in volumetric optical coherence tomography data sets," *https://doi.org/10.1117/1.2772658*, vol. 12, no. 4, p. 041206, Jul. 2007, doi: 10.1117/1.2772658.

[10]  D. Hillmann, "OCT on a chip aims at high-quality retinal imaging.," *Light Sci Appl*, vol. 10, no. 1, pp. 21–21, Jan. 2021, doi: 10.1038/S41377-021-00467-Z.

[11]  S. R. Rufai, "Handheld optical coherence tomography removes barriers to imaging the eyes of young children," *Eye 2021 36:5*, vol. 36, no. 5, pp. 907–908, Jan. 2022, doi: 10.1038/s41433-021-01884-5.

[12]  M. Ruggeri, J. C. Major, C. McKeown, R. W. Knighton, C. A. Puliafito, and S. Jiao, "Retinal Structure of Birds of Prey Revealed by Ultra-High Resolution Spectral-Domain Optical Coherence Tomography," *Invest Ophthalmol Vis Sci*, vol. 51, no. 11, pp. 5789–5795, Nov. 2010, doi: 10.1167/IOVS.10-5633.

[13]     R. J. Zawadzki *et al.*, "Adaptive-optics SLO imaging combined with widefield OCT and SLO enables precise 3D localization of fluorescent cells in the mouse retina," *Biomed Opt Express*, vol. 6, no. 6, p. 2191, Jun. 2015, doi: 10.1364/BOE.6.002191.

[14]     M. J. Booth, "Adaptive optics in microscopy," *Philosophical Transactions of the Royal Society A: Mathematical, Physical and Engineering Sciences*, vol. 365, no. 1861, pp. 2829–2843, Dec. 2007, doi: 10.1098/RSTA.2007.0013.

[15]     Z. Li *et al.*, "Random two-frame interferometry based on deep learning," *Optics Express, Vol. 28, Issue 17, pp. 24747-24760*, vol. 28, no. 17, pp. 24747–24760, Aug. 2020, doi: 10.1364/OE.397904.

[16]     S. A. Boppart, Y.-Z. Liu, P. Pande, and F. A. South, "Automated computational aberration correction method for broadband interferometric imaging techniques," *Optics Letters, Vol. 41, Issue 14, pp. 3324-3327*, vol. 41, no. 14, pp. 3324–3327, Jul. 2016, doi: 10.1364/OL.41.003324.

[17]     "CN208851469U - Liquid lens coherence tomography system and optical coherence tomography devices - Google Patents." Accessed: Jun. 25, 2024. [Online]. Available: https://patents.google.com/patent/CN208851469U/en

[18]     "Liquid lens coherence tomography system and optical coherence tomography devices," Apr. 2018.

[19]     M. Wojtkowski, J. S. Duker, J. G. Fujimoto, V. J. Srinivasan, A. Kowalczyk, and T. H. Ko, "Ultrahigh-resolution, high-speed, Fourier domain optical coherence tomography and methods for dispersion compensation," *Optics Express, Vol. 12, Issue 11, pp. 2404-2422*, vol. 12, no. 11, pp. 2404–2422, May 2004, doi: 10.1364/OPEX.12.002404.

[20]     A. Fercher, C. Hitzenberger, M. Sticker, R. Zawadzki, B. Karamata, and T. Lasser, "Numerical dispersion compensation for Partial Coherence Interferometry and Optical Coherence Tomography," *Optics Express, Vol. 9, Issue 12, pp. 610-615*, vol. 9, no. 12, pp. 610–615, Dec. 2001, doi: 10.1364/OE.9.000610.

[21]     U. Morgner *et al.*, "In vivo ultrahigh-resolution optical coherence tomography," *Optics Letters, Vol. 24, Issue 17, pp. 1221-1223*, vol. 24, no. 17, pp. 1221–1223, Sep. 1999, doi: 10.1364/OL.24.001221.

[22]     A. Baghaie, Z. Yu, and R. M. D'Souza, "State-of-the-art in retinal optical coherence tomography image analysis," *Quant Imaging Med Surg*, vol. 5, no. 4, pp. 60317–60617, Aug. 2015, doi: 10.3978/J.ISSN.2223-4292.2015.07.02.

[23]     S. Bonora *et al.*, "Wavefront correction and high-resolution in vivo OCT imaging with an objective integrated multi-actuator adaptive lens," *Optics Express, Vol. 23, Issue 17, pp. 21931-21941*, vol. 23, no. 17, pp. 21931–21941, Aug. 2015, doi: 10.1364/OE.23.021931.

[24]     "Cascaded Mach-Zehnder Interferometer Sensor on Microring Resonator using Stepper Motor Assisted Optical Wavelength Reuse Ruler." Accessed: Jul. 13, 2023. [Online]. Available: https://opg.optica.org/abstract.cfm?uri=Sensors-2017-SeM2E.4

[25]     K. K. Mishra and S. Ataman, "Optimal phase sensitivity of an unbalanced Mach-Zehnder interferometer," *Phys Rev A  (Coll Park)*, vol. 106, no. 2, p. 023716, Aug. 2022, doi: 10.1103/PHYSREVA.106.023716/FIGURES/14/MEDIUM.

[26]     C. K. Hitzenberger *et al.*, "En-face scanning optical coherence tomography with ultra-high resolution for material investigation," *Optics Express, Vol. 13, Issue 3, pp. 1015-1024*, vol. 13, no. 3, pp. 1015–1024, Feb. 2005, doi: 10.1364/OPEX.13.001015.

[27]     "CN105395162B - 通过电位器控制偏振控制器的方法、装置及oct系统 - Google Patents." Accessed: Jun. 26, 2024. [Online]. Available: https://patents.google.com/patent/CN105395162B/zh

[28] "EP2906985B1 - Compact, low dispersion, and low aberration adaptive optics scanning system - Google Patents." Accessed: Jun. 26, 2024. [Online]. Available: https://patents.google.com/patent/EP2906985B1/en%C2%A0

[29] J. P. Rolland, S. Murali, P. Meemon, W. Kuhn, K. P. Thompson, and K.-S. Lee, "Liquid lens enabled optical coherence microscope with Gabor fusion," *https://doi.org/10.1117/12.871356*, vol. 7652, pp. 781–790, Sep. 2010, doi: 10.1117/12.871356.

[30] O. Puk, C. Dalke, J. Favor, M. H. De Angelis, and J. Graw, "Variations of eye size parameters among different strains of mice," *Mammalian Genome*, vol. 17, no. 8, pp. 851–857, Aug. 2006, doi: 10.1007/S00335-006-0019-5/TABLES/3.

[31] M. I. Seider, R. Y. Lee, D. Wang, M. Pekmezci, T. C. Porco, and S. C. Lin, "Optic disk size variability between African, Asian, white, hispanic, and filipino Americans using heidelberg retinal tomography," *J Glaucoma*, vol. 18, no. 8, pp. 595–600, Oct. 2009, doi: 10.1097/IJG.0B013E3181996F05.

[32] I. Bekerman, P. Gottlieb, and M. Vaiman, "Variations in Eyeball Diameters of the Healthy Adults," *J Ophthalmol*, vol. 2014, 2014, doi: 10.1155/2014/503645.

[33] C. Dysli, V. Enzmann, R. Sznitman, and M. S. Zinkernagel, "Quantitative Analysis of Mouse Retinal Layers Using Automated Segmentation of Spectral Domain Optical Coherence Tomography Images," *Transl Vis Sci Technol*, vol. 4, no. 4, pp. 9–9, Jul. 2015, doi: 10.1167/TVST.4.4.9.

[34] M. Goswami, "AI Pipeline for Accurate Retinal Layer Segmentation Using OCT 3D Images," *Photonics*, vol. 10, no. 3, p. 275, Mar. 2023, doi: 10.3390/PHOTONICS10030275/S1.

[35] A. D. Aguirre, C. Zhou, S.W. Huang, W. Denk, J. G. Fujimoto, and J. Sawinski, "High speed optical coherence microscopy with autofocus adjustment and a miniaturized endoscopic imaging probe," *Optics Express, Vol. 18, Issue 5, pp. 4222-4239*, vol. 18, no. 5, pp. 4222–4239, Mar. 2010, doi: 10.1364/OE.18.004222.

[36] P. Petitt and D. Forciniti, "Cold cataracts: a naturally occurring aqueous two-phase system," *J Chromatogr B Biomed Sci Appl*, vol. 743, no. 1–2, pp. 431–441, Jun. 2000, doi: 10.1016/S0378-4347(00)00220-6.

[37] D. Yadav, D. Chhabra, R. Kumar Garg, A. Ahlawat, and A. Phogat, "Optimization of FDM 3D printing process parameters for multi-material using artificial neural network," in *Materials Today: Proceedings*, Elsevier Ltd, 2020, pp. 1583–1591. doi: 10.1016/j.matpr.2019.11.225.

[38] S. Deswal, R. Narang, and D. Chhabra, "Modeling and parametric optimization of FDM 3D printing process using hybrid techniques for enhancing dimensional preciseness," *International Journal on Interactive Design and Manufacturing*, vol. 13, no. 3, pp. 1197–1214, Sep. 2019, doi: 10.1007/s12008-019-00536-z.

[39] R. Teharia, R. M. Singari, and H. Kumar, "Optimization of process variables for additive manufactured PLA based tensile specimen using taguchi design and artificial neural network (ANN) technique," *Mater Today Proc*, vol. 56, pp. 3426–3432, Jan. 2022, doi: 10.1016/j.matpr.2021.10.376.

[40] S. A. Boppart, Y. Xu, Y.-Z. Liu, and P. S. Carney, "Automated interferometric synthetic aperture microscopy and computational adaptive optics for improved optical coherence tomography," *Applied Optics, Vol. 55, Issue 8, pp. 2034-2041*, vol. 55, no. 8, pp. 2034–2041, Mar. 2016, doi: 10.1364/AO.55.002034.

[41] M. Goswami, "Deep learning models for benign and malign ocular tumor growth estimation," *Computerized Medical Imaging and Graphics*, vol. 93, p. 101986, Oct. 2021, doi: 10.1016/J.COMPMEDIMAG.2021.101986.

[42] L. Zhang, R. Dong, R. J. Zawadzki, and P. Zhang, "Volumetric data analysis enabled spatially resolved optoretinogram to measure the functional signals in the living retina," *J Biophotonics*, vol. 15, no. 3, p. e202100252, Mar. 2022, doi: 10.1002/JBIO.202100252.

[43] T. Sonobe *et al.*, "Comparison between support vector machine and deep learning, machine-learning technologies for detecting epiretinal membrane using 3D-OCT," *Int Ophthalmol*, vol. 39, no. 8, pp. 1871–1877, Aug. 2019, doi: 10.1007/S10792-018-1016-X/FIGURES/4.

[44] M. G. M. Abdolrasol *et al.*, "Artificial neural networks based optimization techniques: A review," *Electronics (Switzerland)*, vol. 10, no. 21. MDPI, Nov. 01, 2021. doi: 10.3390/electronics10212689.

[45] Y. Jian, M. v. Sarunic, R. Ng, M. J. Ju, and D. J. Wahl, "Sensorless adaptive optics multimodal en-face small animal retinal imaging," *Biomedical Optics Express, Vol. 10, Issue 1, pp. 252-267*, vol. 10, no. 1, pp. 252–267, Jan. 2019, doi: 10.1364/BOE.10.000252.

[46] J. L. Tremoleda, A. Kerton, and W. Gsell, "Anaesthesia and physiological monitoring during in vivo imaging of laboratory rodents: Considerations on experimental outcomes and animal welfare," *EJNMMI Res*, vol. 2, no. 1, pp. 1–23, Aug. 2012, doi: 10.1186/2191-219X-2-44/FIGURES/6.

[47] V. Sorrenti, C. Cecchetto, M. Maschietto, S. Fortinguerra, A. Buriani, and S. Vassanelli, "Understanding the Effects of Anesthesia on Cortical Electrophysiological Recordings: A Scoping Review," *International Journal of Molecular Sciences 2021, Vol. 22, Page 1286*, vol. 22, no. 3, p. 1286, Jan. 2021, doi: 10.3390/IJMS22031286.

[48] "Resolution Test Targets." Accessed: Feb. 26, 2023. [Online]. Available: https://www.thorlabs.com/newgrouppage9.cfm?objectgroup_id=4338

[49] C. K. Hitzenberger *et al.*, "Signal averaging improves signal-to-noise in OCT images: But which approach works best, and when?," *Biomedical Optics Express, Vol. 10, Issue 11, pp. 5755-5775*, vol. 10, no. 11, pp. 5755–5775, Nov. 2019, doi: 10.1364/BOE.10.005755.

[50] A. H. Nuttall, "Some Windows with Very Good Sidelobe Behavior," *IEEE Trans Acoust*, vol. 29, no. 1, pp. 84–91, 1981, doi: 10.1109/TASSP.1981.1163506.

[51] "Signal-to-noise ratio - MATLAB snr - MathWorks India." Accessed: Jul. 13, 2023. [Online]. Available: https://in.mathworks.com/help/signal/ref/snr.html

[52] M. Goswami, "AI pipeline for accurate retinal layer segmentation using OCT 3D images," Feb. 2023, doi: 10.48550/arxiv.2302.07806.

[53] A. G. Howard *et al.*, "MobileNets: Efficient Convolutional Neural Networks for Mobile Vision Applications," Apr. 2017, Accessed: Apr. 08, 2024. [Online]. Available: https://arxiv.org/abs/1704.04861v1

[54] F. Valeri *et al.*, "UNet and MobileNet CNN-based model observers for CT protocol optimization: comparative performance evaluation by means of phantom CT images," *Journal of Medical Imaging*, vol. 10, no. Suppl 1, Mar. 2023, doi: 10.1117/1.JMI.10.S1.S11904.

[55] Y. He *et al.*, "Structured layer surface segmentation for retina OCT using fully convolutional regression networks," *Med Image Anal*, vol. 68, p. 101856, Feb. 2021, doi: 10.1016/J.MEDIA.2020.101856.